\documentclass[prd,aps,twocolumn,showpacs,nofootinbib,nobibnotes,superscriptaddress]{revtex4}
\usepackage{epsfig}
\usepackage{amssymb}
\usepackage{graphics}
\usepackage{bm}

\begin{document}

\title{Numerical simulations of the decay of primordial magnetic
turbulence}

\date{\today~~KSUPT-10/1}

\author{Tina Kahniashvili}
\email{tinatin@phys.ksu.edu} \affiliation{McWilliams Center for
Cosmology and Department of Physics, Carnegie Mellon University,
5000 Forbes Ave, Pittsburgh, PA 15213, USA}
\affiliation{Department of Physics, Laurentian University, Ramsey
Lake Road, Sudbury, ON P3E 2C,Canada} \affiliation{Abastumani
Astrophysical Observatory, Ilia State University, 2A Kazbegi Ave,
Tbilisi, GE-0160, Georgia}

\author{Axel Brandenburg}
\email{brandenb@nordita.org} \affiliation{Nordita, AlbaNova
University Center, Roslagstullsbacken 23, 10691 Stockholm, Sweden}
\affiliation{Department of Astronomy, Stockholm University, SE 10691
Stockholm, Sweden}

\author{Alexander G. Tevzadze}
\email{aleko@tevza.org} \affiliation{Abastumani Astrophysical
Observatory, Ilia State University, 2A Kazbegi Ave, Tbilisi,
GE-0160, Georgia} \affiliation{Faculty of Exact and Natural
Sciences, Tbilisi State University, 1 Chavchavadze Ave. Tbilisi,
GE-0128, Georgia}

\author{Bharat Ratra}
\email{ratra@phys.ksu.edu} \affiliation{Department of Physics,
Kansas State University, 116 Cardwell Hall, Manhattan, KS 66506}

\begin{abstract}
We perform direct numerical simulations of forced and freely
decaying 3D magnetohydrodynamic turbulence in order to model
magnetic field evolution during cosmological phase transitions in
the early Universe. Our approach assumes the existence of a magnetic
field generated either by a process during inflation or shortly
thereafter, or by bubble collisions during a phase transition. We
show that the final configuration of the magnetic field depends on
the initial conditions, while the velocity field is nearly
independent of initial conditions.
\end{abstract}

\pacs{98.70.Vc, 98.80.-k}

\maketitle

\section{Introduction}

Astronomical observations show that galaxies have magnetic fields
with a component that is coherent over a large fraction of the
galaxy with field strengths of order $10^{-6}$ Gauss (G); see Refs.\
\cite{Widrow,Vallee,Giovannini2008,beck} and references
therein.\footnote{ See Ref.\ \cite{neronov} and references therein
for a technique that might soon prove useful for measuring a
larger-scale, cosmological, magnetic field.} Modeling the origin of
these fields is a challenging pro\-blem. Here we consider models in
which the seed field is generated in the early Universe, see, e.g.
\cite{Widrow,Giovannini2008,subramanian}. There are a number of such
models and different models result in different magnetic field
structures.

It is well known that quantum-mechanical fluctuations during
inflation can generate energy-density inhomogeneities that seed
observed large-scale structure, see, e.g., Ref.\ \cite{fischler}.
Quantum-mechanical zero-point fluctuations can also generate a seed
magnetic field, provided the relevant abelian gauge field is such
that the lagrangian density is not conformally invariant during
inflation \cite{turner,ratra,bamba}. A seed magnetic field generated
during inflation has a correlation length today that can be very
much larger than the current Hubble radius. A convenient
phenomenological way of breaking conformal invariance during
inflation is to couple the inflaton scalar field $\phi$ to the
vector abelian gauge field by $e^{\alpha\phi} F_{\mu\nu}
F^{\mu\nu}$, where $F_{\mu\nu}$ is the vector field strength tensor
and $\alpha$ is a parameter \cite{ratra}. Depending on the value of
$\alpha$, this can result in a sufficiently large seed magnetic field to
explain the observed galactic magnetic fields. This is a classically
consistent model, with vector field back-reaction during inflation
being negligibly small \cite{ratra,ratralong}. While the abelian
gauge field becomes strongly coupled at early times
\cite{ratra,ratralong,backreaction}, this is not a problem for the
effective, phenomenological, classical model,
\cite{ratra,ratralong,backreaction} and also see
\cite{subramanian,backreaction2}. Of course, just as one does not
reject the effective, phenomenological, classical ``standard''
$\Lambda$CDM cosmological model, one does not reject inflation-based
seed magnetic field generation models for being quantum-mechanically
inconsistent. Rather, much as the case for $\Lambda$CDM, it is of
great interest to understand how such a successful inflation-based
seed magnetic field generation model might arise from a more
fundamental underlying model or theory.

There are a number of other seed field generation models; see Ref.\
\cite{horava} for a discussion of seed field generation in Ho{\v
r}ava-Lifshitz gravity. Yet another interesting possibility is
primordial magnetic field generation during a cosmological phase
transition \cite{pmf}. If the phase transition takes place at late
times the magnetic field correlation length is smaller than the
Hubble radius.

If the magnetic field was generated during or shortly after
inflation (during a pre- or re-heating phase transition), and
survived to the epoch of recombination (the last-scattering
surface), it should leave observable signatures on the cosmic
microwave background (CMB) radiation fluctuations, see Ref.\
\cite{Giovannini2006} for a review and Refs.\ \cite{cmb,faraday} for
more recent discussions.\footnote{A potential advantage of a
primordial seed field that is coherent over scales larger than the
current Hubble radius, as can be generated by inflation, is that it
might be able to explain some of the large-scale oddities of the
observed CMB temperature anisotropy \cite{bernui}, including
potential non-Gaussianity \cite{seshadri}.}

The shape and magnitude of a primordial magnetic field can be
constrained observationally. Magnetic field energy density scales
like radiation (i.e.\ relativistic) energy density.\footnote{The
ratio of the magnetic field energy density $\rho_B$ and the energy
density of radiation $\rho_{\rm rad}$ is constant during
cosmological evolution, if the primordial magnetic is not damped
by MHD (or other) processes and so stays frozen into the plasma.}
During nucleosynthesis, any new form of radiation-like energy
density is constrained observationally to be less than about
$10\%$ of the usual radiation energy density. Hence, agreement
between the big bang nucleosythesis (BBN) model light nuclei
abundance predictions and the observations yields a constraint on
the primordial magnetic field energy density: $\Omega_B h_0^2
\leq 2.4 \times 10^{-6}$ \cite{GrassoR}. Here $\Omega_B$ is the
current value of the magnetic field energy density parameter and
$h_0$ is the current value of the Hubble parameter in units of
100 km s$^{-1}$ Mpc$^{-1}$. Naturally, this limit holds only if
the primordial magnetic field was generated prior to or during
nucleosynthesis.

Other constraints arise through the effects of a primordial magnetic
field on the propagation of CMB photons (Faraday rotation of the CMB
polarization \cite{faraday}, and magnetic field induced scalar,
vector, and tensor modes of CMB anisotropies
\cite{Giovannini2006,cmb}). Available data limit the current value
of a cosmological magnetic field to be less than about $10^{-8}$ G
for a scale-invariant or homogeneous primordial magnetic field.

Another interesting signature of a primordial magnetic field is
relic gravitational waves generated by the anisotropic magnetic
stress \cite{magnet}. The amplitude of the gravitational wave signal
is determined by the efficiency of gravitational wave generation and
the strength of the primordial magnetic field. The efficiency of
gravitational wave production is small \cite{ktr09} due to the small
value of the Newton constant $G$ in the coupling between the
primordial magnetic field and the gravitational waves. However, a
magnetic field that satisfies the BBN limit and that is generated
during a strong-enough first-order phase transition can lead to a
detectable signal for LISA. Therefore, the direct measurement of the
gravitational wave background can lead to an independent limit on a
magnetic field generated in the early Universe \cite{wang}.

On large scales, because of the high conductivity of the plasma in
the early Universe, the magnetic field is treated as a frozen-in
field with its evolution determined by the simple dilution of
magnetic field lines, ${\bf B}({\bf x}, t) \propto {\bf B}({\bf
x})/ a^2(t)$, where $t$ is the physical cosmic time and $a(t)$ is
the scale factor. In general, however, the evolution of a
primordial magnetic field is a complex process influenced by MHD
as well as by cosmological dynamics \cite{banerjee}. In
particular, the presence of a magnetic field can dramatically
affect the behavior of primordial turbulence (for example,
turbulence associated with phase transition bubble motions)
\cite{axel-1,axel-2,caprini2009}. Also, the presence of a magnetic
field might itself lead to the development of turbulent motions,
see Refs.\ \cite{B, Verma, axel} and Sec.~III below.

In this paper we study the evolution of a primordial magnetic
field that is coupled, through the MHD equations, with the fluid,
during a cosmological phase transition. We consider two different
possibilities: (i) when the magnetic field energy has been
injected in the plasma with no initial vorticity perturbations
(no turbulent motions) at a given typical scale; and, (ii) when
vorticity perturbations are present during the phase transition
and couple to the magnetic field with a given spectrum. Both
these assumptions can be justified. The magnetic field might be
generated prior to the phase transition, with the generation
mechanism not requiring turbulent motion. On the other hand, the
presence of initial turbulent motions can also be justified since
bubble nucleation, expansion, and collisions can lead to
primordial kinetic turbulence \cite{kos1}. Of course, in reality
both magnetic field and kinetic turbulence generation can take
place together, while in our approach we assume the a priori
presence of magnetic energy and/or kinetic energy during the
phase transition.

A causal, phenomenological description of MHD on large scales during
the cosmological phase transition motivates a white noise spectrum
for the magnetic spectral energy density, $E_M(k) \propto k^2$,
where $k$ is the wavenumber \cite{hogan}. A steeper
Batchelor spectrum with $E_M(k) \propto k^4$ has been claimed to
follow from cosmological causality and the divergenceless condition
of the magnetic field \cite{Durrer}. To understand the evolution of
a cosmological magnetic field, and to be able to make observable
predictions in this model, it is important to resolve this impasse.
One would like to know whether the final large-scale magnetic
energy spectrum evolves to something steeper than white
noise, or to a spectrum closer to the Batchelor one.

In fact, the MHD process itself might establish a different
spectrum: The possibility of generating a random magnetic field
from isotropic turbulence was first proposed by Batchelor
\cite{batchelor}. He invoked the imperfect analogy between
vorticity and magnetic fields that should generally imply a large
scale distribution of the magnetic energy similar to the kinetic
energy of turbulent motions $E_K(k) \propto k^4$
\cite{batchelor2}. Later, Kazantsev \cite{kazantzev} was able to
rigorously establish the possibility of small-scale dynamo
action. He assumed that the velocity field varied only on large
scales and found that weak magnetic fields are amplified mainly
on the resistive scale. Initially, this was thought to be
applicable only to turbulence at large magnetic Prandtl number
\cite{Scheko}, where the viscous cutoff scale is much larger than
the resistive cutoff scale. Kazantsev found that the magnetic
energy spectrum increases with wavenumber like $k^{3/2}$, which
is slightly shallower than the white noise spectrum $k^2$. The
emergence of a Kazantsev spectrum turned out to be much more
ubiquitous and not only applicable at large magnetic Prandtl
number. Simulations at magnetic Prandtl numbers of order unity
also clearly showed the Kazantsev $k^{3/2}$ spectrum
\cite{axel-2}. Such a spectrum could be of interest for
primordial magnetic field evolution, because it implies somewhat
larger power at large scales than white noise.

The main goal of our paper is to determine, through MHD modeling,
the evolution of the magnetic energy spectrum. To keep our study as
general as possible we do not make any assumption about the physical
process leading to primordial magnetic field generation. Also, we do
not address the phase transition physics itself, keeping the total
magnetic field energy density $\rho_B$ as a free parameter whose
maximal value is fixed by the BBN bound. Obviously, if the magnetic
field is generated during a phase transition, $\rho_B$ will depend
sensitively on the amount of latent heat that is transformed to
magnetic energy, i.e.\ on the efficiency of the magnetic field
generation process. One important issue addressed here is to
determine the spectral shape of the magnetic field at large scales,
assuming that magnetic energy and vorticity perturbations are
closely coupled during the phase transition. Another important
question is related to the duration of MHD turbulence and how long
it takes to reach equipartition between kinetic and magnetic energy
densities in the primordial plasma.

The structure of this paper is as follows. In Sec.\ II we describe
the model, defining the main model parameters (Sec.\ IIA),
characterizing the magnetic field spectrum (Sec.\ IIB), and
formulating initial conditions (Sec.\ IIC). In Sec.\ III we present
results from direct 3D MHD simulations performed using the {\sc
Pencil Code} \cite{pencil}. We conclude in Sec.\ IV. We employ
natural units with $\hbar = 1 = c$ and gaussian units for
electromagnetic quantities. To properly account for the expansion of
the Universe we use comoving quantities with conformal time $t$ that
is related to physical time $t_{\rm phys}$ as $t \propto t_{\rm
phys}^{1/2}$ during the radiation dominated epoch.

\section{Model Description}

We assume that magnetic energy is generated at the electroweak or
QCD phase transition, or during inflation. In the first case
magnetic energy is explicitly injected into the fluid on small
length scales, smaller than the Hubble length at the moment of the
phase transition. If the magnetic field originated at an earlier
epoch of inflation the length scale at which the magnetic field
interacts with the fluid is set by the characteristic length scale
of the system, and is again smaller than the Hubble radius. In the
absence of magnetic or kinetic helicity we do not expect an inverse
cascade (i.e., energy flow from smaller to larger length scales).

\subsection{Phase transition characteristics}

To model magnetic field evolution one needs to know the physical
conditions during the phase transition. First, generation of
turbulence requires a first-order phase transition so that at the
critical temperature of the phase transition bubbles of the new
vacuum nucleate within the false vacuum. Bubble collisions then
generate turbulent motions. The standard electroweak model does
not have a first-order phase transition~\cite{klrs}, and cannot
account for baryogenesis. However, modifications of the standard
model, such as the minimal supersymmetric standard model
(MSSM)~\cite{r90}, result in first-order electroweak phase
transitions and can account for baryogenesis~\cite{laine}. Also,
recent lattice QCD computations \cite{qcd} have not yet excluded
that the QCD phase transition is a first-order one, with bubble
nucleation and collisions.

The main parameter characterizing turbulent motions is the r.m.s.\
velocity $v_0$ which determines the kinetic energy density of the
turbulence.  Obviously, $v_0$ depends sensitively on the phase
transition physics, and in particular on the phase transition bubble
wall expansion velocity $v_b$, \cite{vb}. To model the development
of turbulent motions during the phase transition we adapt earlier
analytical or semi-analytical results \cite{vb,nicolis}. The first
question that must be answered is whether the phase transition is
first order, and, if so, what fraction of total available vacuum
energy is transformed into kinetic energy of the bubbles. The
r.m.s.\ velocity of the turbulent motions can be approximated as
(see Ref.\ \cite{nicolis})
\begin{equation}
v_0  =  \sqrt{\frac{3\kappa\alpha}{4 + 3\kappa \alpha}}. \label{v0}
\end{equation}
Here $\alpha = \rho_{\rm vac} / \rho_{\rm thermal}$ is the ratio of
the vacuum energy density associated with the phase transition to
the thermal energy density of the Universe at the time ($\alpha$
characterizes the strength of the phase transition), and the
efficiency factor $\kappa$ is the fraction of the available vacuum
energy that goes into the kinetic energy of the expanding bubble
walls (as opposed to thermal energy).

On the other hand, the primordial magnetic field is characterized by
the r.m.s.\ Alfv\'en velocity given by
\begin{equation}
v_A =  \frac{B}{\sqrt{4\pi {\rm w_{\rm rad}}}}\; =
\sqrt{\frac{3\rho_B}{2 \rho_{\rm rad}}} \simeq 7.65 \times 10^2
\sqrt{\Omega_B}. \label{vA}
\end{equation}
Here ${\rm w}_{\rm rad}= 4\rho_{\rm rad}/3$ (with $\rho_{\rm rad}
\simeq \rho_{\rm thermal}$) is the radiation enthalpy of the
relativistic fluid and we have used $\Omega_{\rm rad} h_0^2= 2.56
\times 10^{-5}$ for a present-day CMB temperature $T_0=2.74$ K. At
temperature $T_\star$ of the phase transition we have $\rho_{\rm
rad}(T_\star) = {\pi^2}g_* (T_*)^4/30 $, where $g_\star$ is the
number of relativistic degrees of freedom at temperature $T_\star$.
The r.m.s.\ Alfv\'en velocity does not depend on $T_\star$, but it
is weakly dependent on $g_\star$,
\begin{equation}
v_A \simeq 4 \times 10^{-4} \left( \frac{B}{10^{-9}{\rm
G}}\right)\left(\frac{g_\star}{100}\right)^{-1/6}. \label{va1}
\end{equation}
Assuming equipartition between kinetic and magnetic energy
densities, one has $v_A \simeq v_0$. While the equipartition
condition can be justified by MHD dimensional analysis \cite{B}, we
explicitly show that it holds by performing a 3D direct numerical
simulation of a primordial magnetic field coupled to fluid motions;
see Sec.\ III. Also, partial equipartition, $v_0 \sim 0.8 v_A$, is
reached in the case where the magnetic field was generated first at
the injection scale, which then led to rapid growth of vorticity
perturbations in the initially no-turbulent plasma.

The bubble wall expansion sets the maximal size of the bubble, which
we associate with the size of the largest turbulent eddy, $l_0 = v_b
\beta^{-1}$, where $\beta$ is a parameter that characterizes the
duration of the phase transition. In particular, $\beta $ can be obtained
from the bubble nucleation rate \cite{kos1}. The Hubble time
$H_\star^{-1}$ at the phase transition is another characteristic
time, and it sets the ``causality" horizon. At this point it is
useful to define the parameter $\gamma [= l_0 H_\star  = v_b
(\beta/H_\star)^{-1}]$ which determines how many maximal-sized
bubbles are within the Hubble radius, $N \approx \gamma^3$.

\subsection{Magnetic field spectrum}

If a primordial magnetic field is randomly oriented and its mean
value vanishes, i.e.\ $\langle {\bf B(x)} \rangle =0$, it is
conveniently described statistically in terms of the $n$-point field
correlation functions. If the field is isotropic with a Gaussian
distribution, the magnetic field characteristics are completely
determined by the two-point correlation function $\langle {\bf
B}({\bf x +r}, t+\tau) {\bf B}({\bf x}, t) \rangle$. To construct
this main characteristic function we need to know the spatial
distribution (i.e., the correlation length) and the temporal
evolution of the magnetic field.

When considering a causally generated primordial magnetic field, its
maximal co-moving correlation length $\xi_{\rm max}$ is set by the
co-moving Hubble radius $\lambda_H (= H_\star^{-1} a_0/a_\star)$ at
the moment of generation (here $a_0$ and $a_\star$ are the scale
factors today and at magnetic field generation, respectively).
Causality implies that
\begin{equation}
\xi_{\rm max} \leq \lambda_{H} = 5.8 \times 10^{-10}~{\rm
Mpc}\left(\frac{100\,{\rm GeV}}{T_\star}\right)
\left(\frac{100}{g_\star}\right)^{{1}/{6}}, \label{lambda-max}
\end{equation}
where the temperature $T_\star$ corresponds to the energy scale at
field generation. For the electroweak phase transition, $T_\star$ is
related to the Higgs mass $M_H$ through $T_\star \simeq (1.2 \pm
0.2) M_H$, so we parameterize the $T_\star$ dependence by
normalizing to a temperature of $100$ GeV. Also, for the electroweak
phase transition $g_\star \simeq 100$, but the dependence on
$g_\star$ is much weaker than the $T_\star$ dependence.

Another way to determine the magnetic field correlation length is to
use the magnetic energy spectrum\footnote{In what follows we use
\cite{my75}
\begin{equation}
\langle B_i^* ({\bf k}, t) B_j({\bf k'}, t+\tau) \rangle =
\delta({\bf k} -{\bf k'}) \, F_{ij}^M\!({\bf k},t) \,
f[\eta(k),\tau], \label{2-point}
\end{equation}
where
\begin{equation}
8\pi k^2 F_{ij}^M\!({\bf k},\tau) =  2 P_{ij}({\bf k}) E_M(k,t) + i
\varepsilon_{ijl} {k_l} H_M(k,t) \, . \label{eq:4.1}
\end{equation}
Here $P_{ij}({\bf k}) = \delta_{ij} - {k_i k_j}/{k^2}$ is the projection
operator, $\delta_{ij}$ is the Kronecker delta, $k = |{\bf k}|$,
$\varepsilon_{ijl}$ is the totally antisymmetric tensor, and
$\eta(k)$ is an autocorrelation function that determines the
characteristic function $f[\eta(k),\tau]$ describing the temporal
decorrelation of turbulent fluctuations. The scalar function
$H_M(k,t)$ is the magnetic helicity spectrum.  All configurations of
the helical magnetic field must satisfy the ``realizability
condition'' \cite{B,Verma}, $|\int_0^\infty dk {H}_M(k, t)| \leq 2
\xi_M(t) {\mathcal E}_M(t)$.} $E_M(k, t)$, defining the correlation
length by $\xi_M (t) = \left[\int_0^\infty {\rm d}k\, k^{-1}
E_M(k,t)\right]/\int_0^\infty {\rm d}k\, E_M(k,t)$.

In most models of magnetic field generation during a phase
transition \cite{pmf}, the magnetic field correlation length is
determined by the phase transition's bubble sizes. In this case a
characteristic magnetic field correlation length is assumed to be
determined by the largest bubble, $\lambda_0 \simeq \gamma
\lambda_H$. This simple dimensional description implies that the
magnetic energy density $\rho_B$ is redistributed through MHD
processes and establishes a magnetic spectrum with spectral
energy density measured today that is given by \cite{ktr09}
\begin{eqnarray}
E_M({k} ) &\leq &  \frac{5.2
(\alpha+1)}{3\alpha+5}~\left(\frac{100\ {\rm GeV}}{T_\star}\right)
\left(\frac{100}{g_\star}\right)^{1/2} \gamma \nonumber \\
&&\times \frac{(10^{-9}~{\rm G})^2} {{\rm pc }^{-1}}
 \left \{
\begin{array}{c}
{\bar k}^{\alpha} ~~~~~ {\rm if} ~~~ {\bar k}<1 \\
{\bar k}^{-5/3} ~~ {\rm if} ~~~ {\bar k} > 1
\end{array}
\right\} . \label{em}
\end{eqnarray}
Here, ${\bar k}=k/k_0$ and ${\bar k}_D = k_D/k_0$ with $k_0 =
2\pi/\lambda_0$, and $k_D=2\pi/\lambda_D$ is the damping
wavenumber determined through the viscosity-driven dissipation of
the magnetic field. In the case of stationary non-helical
Kolmogorov turbulence the Reynolds number determines the damping
scale $k_D$ as  ${\rm Re} = (k_D/k_0)^{4/3}$. Note that the
Reynolds number is high enough in the early Universe to ensure
the presence of a wide inertial (turbulent) range,
$k_0<k<k_D$.\footnote{References \cite{caprini2009} have
recently analytically estimated the Reynolds number in the early
Universe at the scale of energy injection of the turbulence and
magnetic field.} The large-scale behavior of the magnetic field is
determined by the parameter $\alpha$. The scale-invariant
spectrum corresponds to $\alpha =-1$ \cite{ratra}, the Kazantsev
spectrum has $\alpha=3/2$ \cite{kazantzev}, the white noise
spectrum corresponds to $\alpha =2$ \cite{hogan}, and the steep
Batchelor spectrum has $\alpha=4$ \cite{batchelor}. Equation
(\ref{em}) does not account for any damping of magnetic
energy and can be viewed as the BBN bound imposed at the moment
of the establishment of the magnetic spectrum; see Fig.\ 1.

\begin{figure}
\begin{center}
\includegraphics[width=\columnwidth]{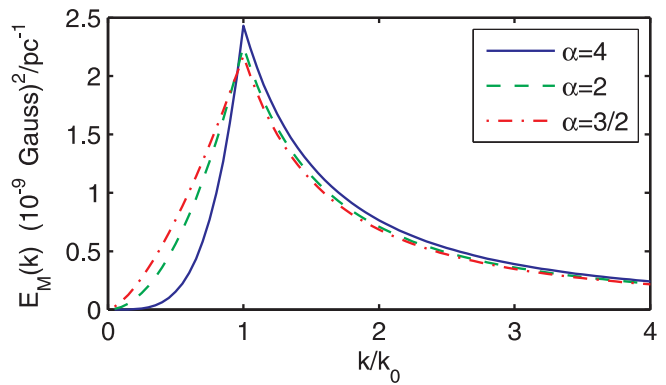}
\end{center}
\caption{The maximal allowed value $E_M(k)$ of the magnetic field
generated during the EW phase transition with $T_\star=100$ Gev,
$g_\star=100$, and $\gamma=0.01$.} \label{EM}
\end{figure}

\subsection{MHD formalism and initial conditions}

As discussed in Ref.~\cite{enqvist}, the usual relativistic MHD
equations are identical to the MHD equations in an expanding
Universe with zero spatial curvature when physical quantities are
replaced by their co-moving counterparts and conformal time
${\eta}$ is used in place of physical time. Based on this fact,
we perform direct numerical simulations of MHD turbulence in the
expanding Universe using the usual MHD equations for an
ultrarelativistic equation of state, but nonrelativistic bulk
motions.

The characteristic length-scale of the problem is set by the phase
transition bubble size $\gamma \lambda_H$. The typical time-scale is
the eddy turn-over time. Assuming that turbulent eddies correspond
to phase transition bubbles, the {\it physical} turn-over time is
$\tau_0 \sim (v_b/v_0) (\beta/H_\star)^{-1}$, where $v_0 = \langle
V^2 \rangle^{1/2}$.

Another characteristic of the initial stage is the amount of
magnetic and kinetic energies present during the phase transition,
i.e.\ the initial Alfv\'en and turbulent r.m.s.\ velocities. As
discussed above, both quantities are sensitively dependent on the
available vacuum energy that is converted to magnetic energy (if the
magnetic field was generated during the phase transition) and/or the
kinetic energy of the turbulent motions. In our analysis below we
assume that about $10\%$ of the vacuum energy is in the form of
initial magnetic energy, which corresponds to $v_{A, {\rm in}}
\simeq 0.3$.

We perform the numerical simulations of magnetic field evolution
in two stages. During the first stage, we model \emph{forced} MHD
turbulence with injection of energy at fixed wavenumber. This type
of driving force is supposed to mimic the action of bubble-induced
external forces in MHD turbulence during the phase transition. The
simulation of forced MHD turbulence is carried out before
equipartition is reached, i.e., before $v_0 \sim v_A$.  After
equipartition is reached we switch off the driving force and allow
\emph{free decay} of the turbulent state.

We use two different types of external forcing during the first
stage of the simulations. These are injecting magnetic or kinetic
energy at a given scale associated with the turbulent eddy size.
These cases differ in the initial magnetic field configuration.
Injection of the magnetic energy in the flow is achieved by using a
delta function spectral energy density function for the magnetic
field. Injection of the kinetic energy is achieved by using a delta
function spectral energy density function for the velocity field. In
the latter case the spectral energy distribution of the initial
magnetic field is of Batchelor's type ($E_B \propto k^4$) and its
amplitude is close to that of the kinetic energy density (the
equipartition condition).

We also define the spectrum of the velocity field $E_K(k,t)$ and the
total energy density ${\mathcal E}_K(t) = \int_0^\infty E_K(k,t)\,
{\rm d}k$. One of our main goals is to determine whether the
presence of magnetic fields in the early Universe (e.g.\ generated
prior to a phase transition) can lead to strong turbulence.

\section{Numerical Simulations}

Numerical simulations of the magnetic field evolution were performed
using the {\sc Pencil Code} \cite{bd2002,b2003,axel-2,dsb2006}; see
\cite{pencil} for the website.

We perform all simulations using co-moving quantities and conformal
time. For simplicity we work with dimensionless quantities, and use
$k_1=k_0/30$ as our wavenumber unit. The chosen box size covers a
wavenumber range from $k_0/30$ up to $4.3\times k_0$, and the
maximal length scale considered corresponds to $\lambda_1= 30 \gamma
\lambda_H$, which is still within the electroweak phase transition
Hubble scale ($\gamma_{\rm EW} \leq 0.01$). In the case of a QCD
phase transition for an extremely large QCD bubble velocity, $v_b
\rightarrow 1$, the scale can exceed the QCD phase transition Hubble
scale by a factor of 2 ($\gamma_{\rm QCD} \leq 0.15$). Of course, in
the simulations we have to use a relatively large value for the
dissipation wavenumber, which is $k_D \simeq 2 k_0$ at the end of
the simulation. This high value is a consequence of choosing a
constant viscosity that must be large enough so that it can
also cope with the initially much larger value of the energy
dissipation rate. Such high values are obviously not realistic for
the early Universe where the Reynolds number is extremely high.
However, we motivate this choice by the fact that we are mainly
interested in the evolution of the magnetic field outside the
inertial range for $k<k_0$. The time unit in our simulations is set
by the computational box size and sound speed, i.e.,
$t_1=\sqrt{3}/k_1$.

\begin{figure*}[t]
\includegraphics[width=\textwidth]{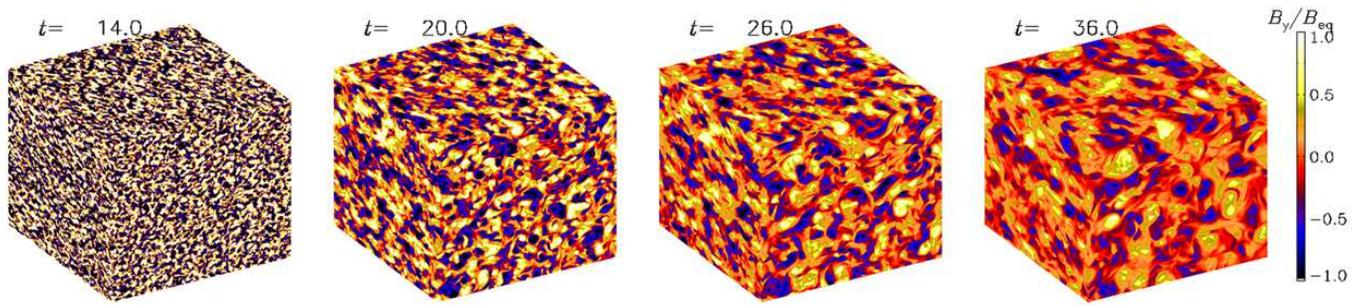}
\caption{Evolution of the turbulent magnetic field after turning off
the forcing at time $t=14\,t_1$. The $B_y$ component is shown on the
periphery of the computational domain.} \label{B_xyz}
\end{figure*}

As is well known, free decay of MHD turbulence implies an increase
of the magnetic eddy size with decreasing magnetic energy
density. Figure \ref{B_xyz} illustrates this fact in our
simulations. We display the $y$ component of the magnetic field on
the periphery of the domain during stages of its evolution after
equipartition has been reached and driving was switched off, in the
case when the magnetic energy was injected at some typical scale.

The characteristic time scales for the qualitative changes in the
magnetic field distribution is approximately equal to $20t_1 \simeq
165 \gamma \lambda_H$ -- a value that exceeds slightly the Hubble
time-scale, $\lambda_H/c$, for the electroweak phase transition.

\subsection{Development toward equipartition}

\begin{figure}[t!]
\includegraphics[width=\columnwidth]{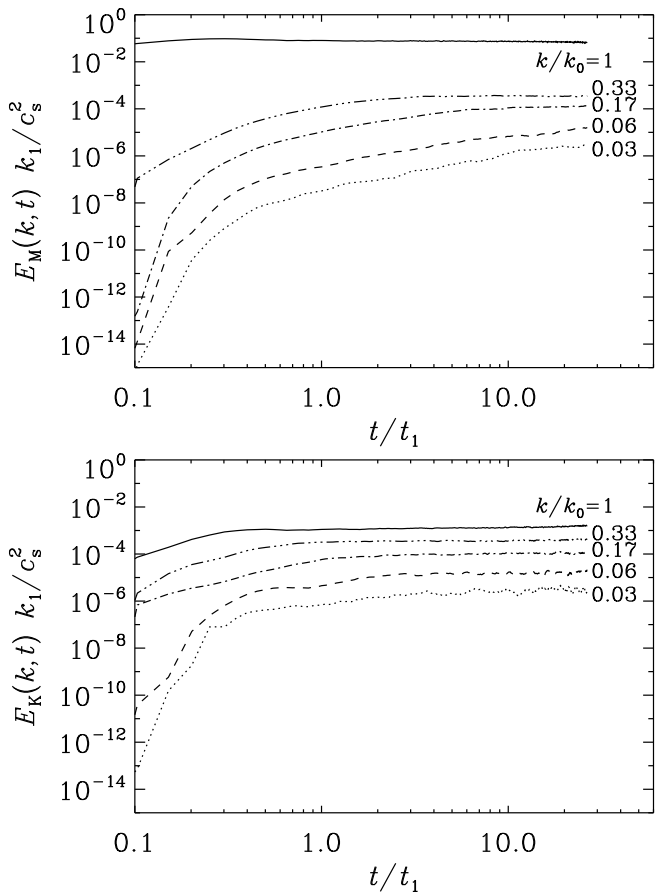}
\caption{Evolution of the magnetic energy spectral density
$E_M(k,t)$ (upper panel) and kinetic energy spectral density
$E_K(k,t)$ (lower panel) for different values of $k/k_0=$ 0.03,
0.06, 0.17, 0.33, and 1. The spectra are normalized such that $\int
E_{\rm M}(k,t)dk=\langle{\bm u}^2\rangle/2$.
}\label{pfm_256d}\end{figure}

\begin{figure}[t]
\includegraphics[width=\columnwidth]{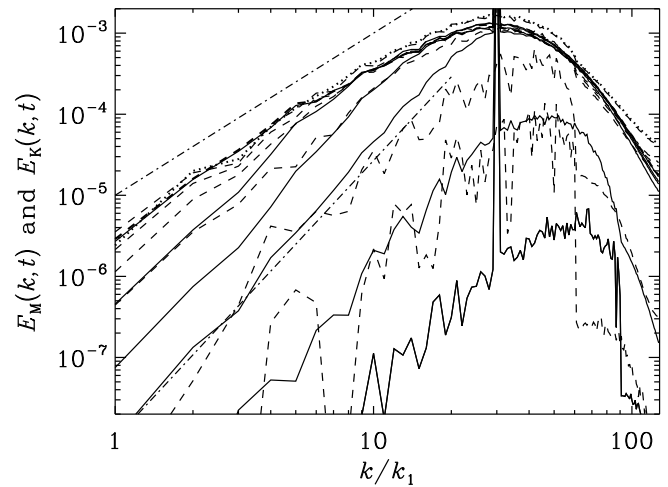}
\caption{The initial temporal evolution of magnetic energy spectra
$E_M(k)$ [solid lines at $t/t_1=0.1$, 0.2, 0.5, 1.5, 5, and 25, with
(smoothed) $E_M(k)$ at $k = 10 k_1$ increasing as $t$ increases] are
shown before the field reaches equipartition with the kinetic energy
density. For comparison, kinetic energy spectra are shown for the
same times (dashed lines). Thick lines (solid and dashed) indicate
the last time, $25\,t_1$. Straight dash-dotted lines have slopes 2
and 4. The box turnover time is $15\,t_1$ and is reached only for
the last time shown. } \label{Evolution}
\end{figure}

We first examine how long it takes for forced MHD turbulence to
establish equipartition between the magnetic and kinetic energy
densities, i.e.\ where $v_A\simeq v_0$. Of course, the evolution of
the magnetic energy spectrum is strongly scale dependent. We first
address the case when the magnetic field is ``created'' at a typical
scale $k_0^{-1}$, corresponding to the largest bubble. The Alfv\'en
velocity of this field is close to the maximal value allowed by the
BBN bound, i.e.\ $v_{A, {\rm in}} \simeq 0.3$. In this case $10\%$
of the available vacuum energy has been somehow transformed into
magnetic energy before or during the phase transition. We perform
our simulations with zero initial velocity perturbations. Such
initial conditions may apply to the generation of turbulent motions
during a phase transition via MHD processes, while the magnetic
field was generated prior to that, for example through quantum
fluctuations \cite{ratra,bamba}. In this case the evolution of
magnetic and kinetic energy spectra during the first stages of MHD
coupling are shown in Fig.~\ref{pfm_256d}. Taking into account that
the largest bubble size must be a typical length scale for
turbulence, our initial conditions imply that at $k=k_0$
equipartition is reached almost instantaneously. Larger scales need
substantially longer times to establish equipartition. The initial
evolution of magnetic energy spectra is shown in
Fig.~\ref{Evolution}.

\begin{figure}[t]
\includegraphics[width=\columnwidth]{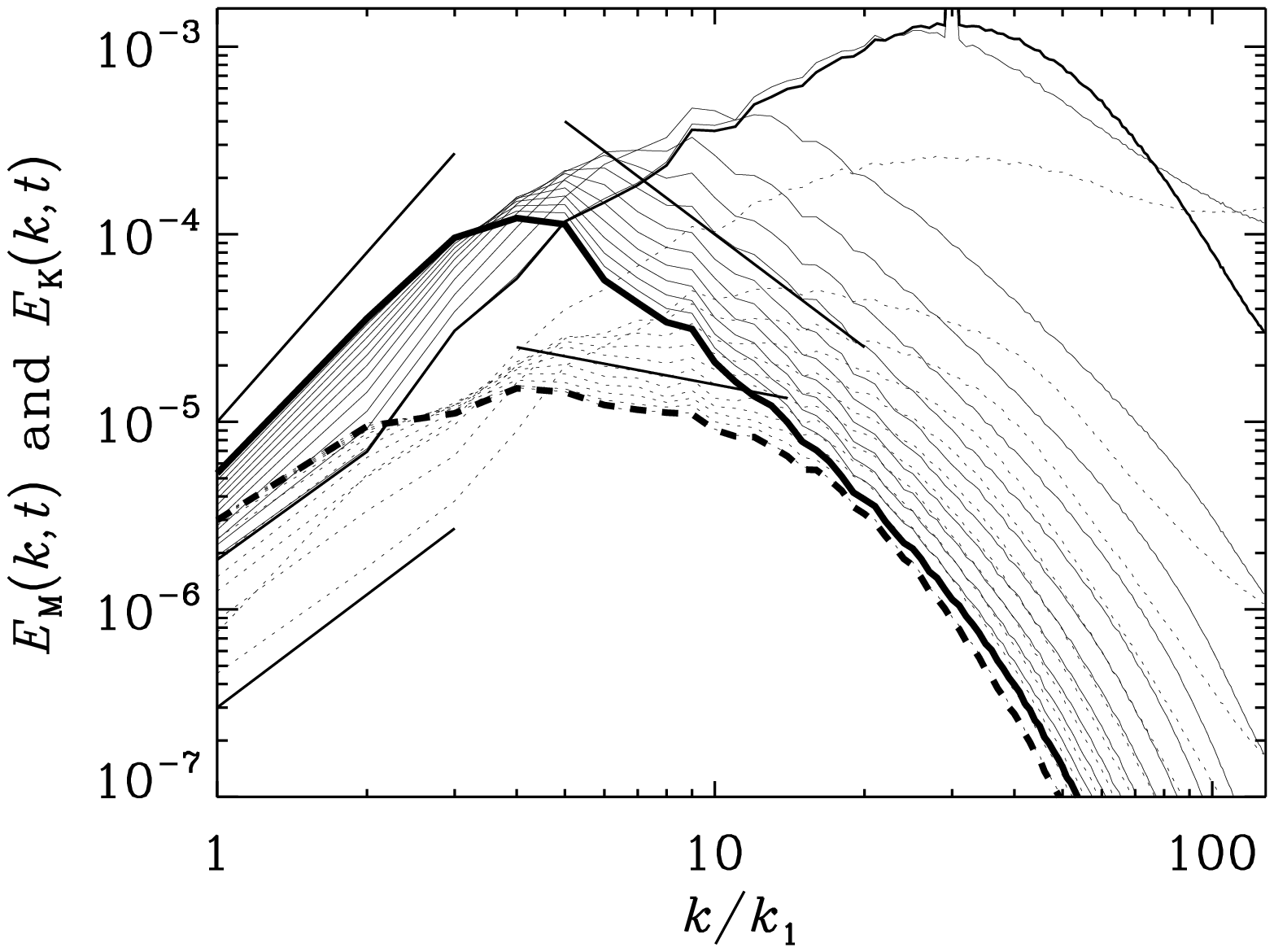}
\caption{Magnetic (solid) and kinetic (dashed) energy spectra in 12
regular time intervals of $4\,t_1$ after having turned off the
forcing, with (smoothed) spectra at $k = 50 k_1$ decreasing as $t$
increases. $\nu=\eta=10^{-4}$ in units of $(k_1^2t_1)^{-1}$. The
straight lines have slopes $3$, $2$, $-2$, and $-1/2$, with the
first two near $k = k_1$ and the last two near $k = 10 k_1$.
Thickest lines (solid and dashed) indicate the last time, which is
$44\,t_1$ since turning off the forcing. The intermediate thickness
solid line, the highest or almost highest line for $k/k_1 > 10$, is
the initial magnetic spectrum for this
computation.}\label{pkt_256d_noforce}
\end{figure}

\begin{figure}[t]
\includegraphics[width=\columnwidth]{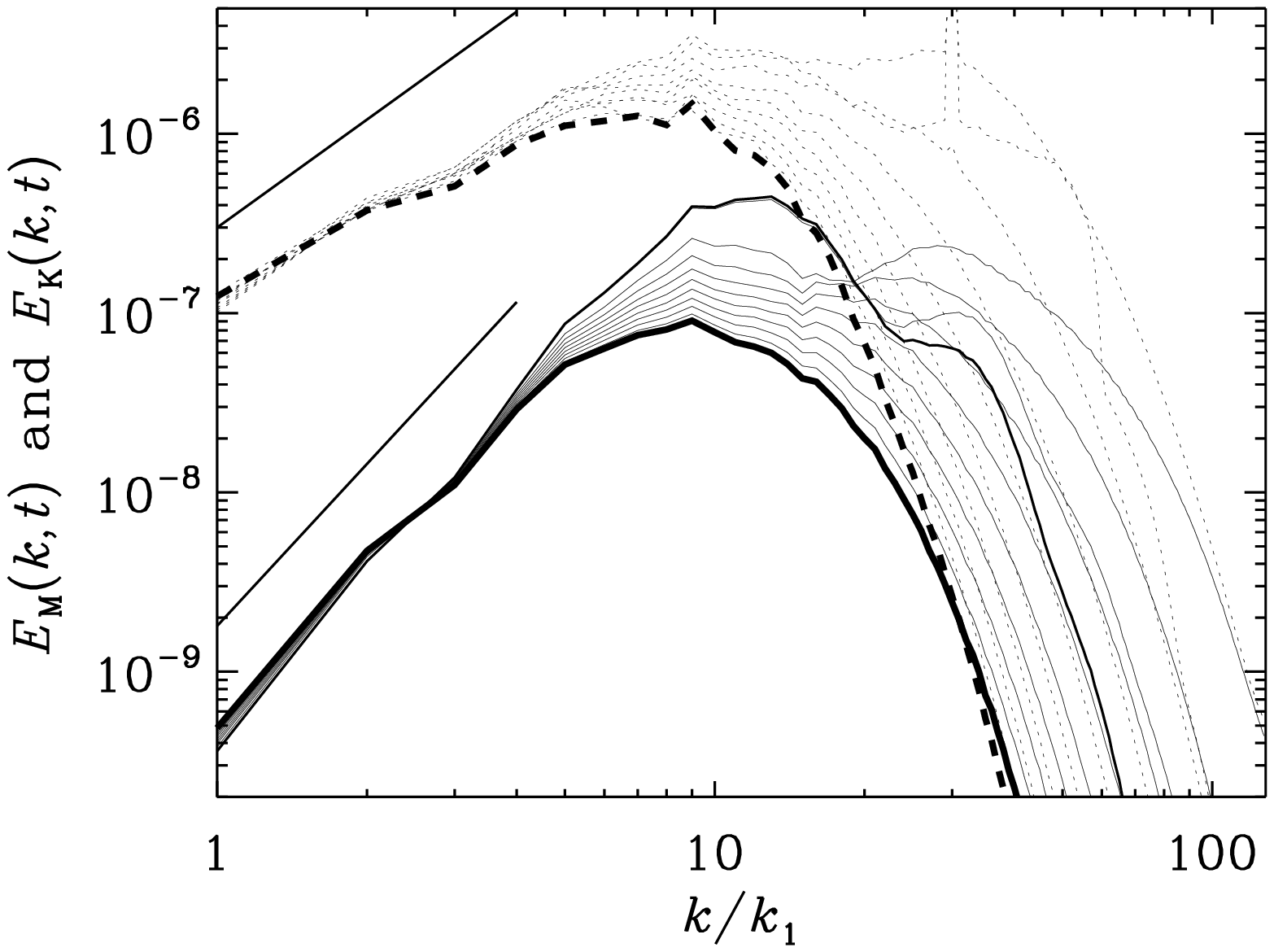}
\caption{Same as Fig.~\ref{pkt_256d_noforce}, but for a case where
the initial magnetic field had a $k^4$ spectrum close to
equipartition with the velocity field, and then the forcing was
turned off. Results are shown for nine times at intervals of
$6\,t_1$. $\nu=\eta=10^{-4}$ in units of $(k_1^2t_1)^{-1}$. The
straight lines have slopes 2 and 3. Thickest lines (solid and
dashed) indicate the last time, which is $48\,t_1$ since turning off
the forcing. The intermediate thickness solid line, the highest
solid line for $5 < k/k_1 < 10$, is the initial magnetic spectrum
for this computation. }\label{pkt_u256d2_noforce2}
\end{figure}

In our phenomenological description it has been assumed
\cite{kos3,nicolis} that bubble collisions and nucleation lead to
the development of turbulent motions, and the kinetic energy
spectrum has been approximated as Kolmogorov-like in the inertial
range ($E_K \propto k^{-5/3}$) and by a white noise spectrum ($E_K
\propto k^2$) for $k<k_0$. As noted above, our simulations cannot
adequately describe the inertial range and thus we cannot expect to
see a Kolmogorov-like spectrum. On the other hand, the large scale
configuration of the velocity field is well approximated by white
noise; see Figs.~\ref{Evolution}, \ref{pkt_256d_noforce}, and
\ref{pkt_u256d2_noforce2}. This agrees with what is predicted in the
phenomenological approach.

We note that the kinetic energy spectrum grows faster right after
the driving is switched on. In particular, after $t_1$, the r.m.s.\
velocity increases from zero to $0.17$ while approaching $0.21$
after $20t_1$. At $v_0 \sim 0.8 v_A$ we assume that equipartition is
reached and, by switching off the driving, we allow the turbulence
to enter the second, free-decay stage.

\subsection{Free decay of MHD turbulence}

The free decay of turbulence is shown in
Figs.~\ref{pkt_256d_noforce} and \ref{pkt_u256d2_noforce2} for two
different initial configurations of the magnetic field. In both
cases we perform MHD simulations after the driving force was
switched off and free decay occurred.

The first case corresponds to the process described above, so it was
preceded by magnetic field injection into the turbulent plasma at
the scale of the largest bubbles or turbulent eddies. The field was
then allowed to decay, leading to the development of near
equipartition between the magnetic and turbulent energies. The
second case corresponds to an initial configuration of a random
magnetic field with a $k^4$ spectrum. In contrast to the first case,
the process here can be roughly described as turbulent kinetic
energy injection into the magnetized plasma. In this case we
approximate the magnetic field spectrum at large scales by the steep
Batchelor spectrum ($E_M\propto k^4$).  Of course, the real
situation during the phase transition is somewhere in between. If
the magnetic field was generated during the phase transition through
bubble collisions \cite{pmf} the same process of bubble collisions
leads to the generation of turbulent motions (vorticity
perturbations). Thus, strictly speaking, we cannot split the
turbulent motions and magnetic field generation and evolution. If
the magnetic field was created before the phase transition, it
affects the bubble collisions (locally inserting a preferred
direction). As a result, the process of generation of turbulent
motions is affected and this backreacts onto the magnetic field
configuration itself.

The first case --- injection of magnetic energy into the turbulent
plasma --- provides a suitable setting for magnetic field generation
through the mechanism described in Ref.~\cite{kisslinger}, where the
correlation length of the magnetic field is naturally set by the
size of the phase transition bubbles. The second case with an
established magnetic field spectrum is more appropriate for the
causal mechanisms discussed in Ref.~\cite{Durrer}.

Figures~\ref{pkt_256d_noforce} and \ref{pkt_u256d2_noforce2} show a
distinctive difference between the turbulent states developed after
the injection of the magnetic energy through a single mode magnetic
field, and the injection of kinetic energy in the existing magnetic
field with smooth magnetic spectral energy density distribution. A
major difference is revealed in the spectral distribution of
magnetic and velocity fields at large scales. At large scales, the
spectral slopes are approximately $2$ for kinetic energy and around
$3$ for magnetic energy. Therewith, the spectral distribution of
kinetic energy shows less sensitivity to initial conditions and
develops a shape close to white noise with $E_k \propto k^2$. The
final configuration of the large-scale magnetic field slightly
differs as a consequence of the initial conditions. In the case of
magnetic energy injection (Fig.~\ref{pkt_256d_noforce}) the spectral
slope is shallower than 3, and tends to establish a white noise
spectrum, while the case of kinetic energy injection
(Fig.~\ref{pkt_u256d2_noforce2}) most probably results in the steep
Batchelor spectrum.

In all cases, throughout the free decay stage, the peak of the
magnetic field spectral energy density drifts to smaller
wavenumbers as $k_{\rm peak}(t) \propto t^{-1/2}$. At the same
time magnetic power decreases as $E_M(k,t) \propto t^{-1}$.
Accounting for these scalings and noting that we use conformal
time, the temporal scaling in the case of an expanding Universe
is somewhat slower than that in the case of laboratory grid
turbulence where $E_K \propto t_{\rm phys}^{-n}$ with exponents
between $n=1.13$ \cite{krogstad} and $1.25$ \cite{KCM03}.

Simulations show that equipartition between kinetic and magnetic
energies is sustained throughout the free decay stage of the
turbulence. Therewith, at any given point in time, the kinetic
energy spectrum peaks at the same wave-number as the magnetic energy
spectrum. However, it seems that properties of the turbulent states
developed through free decay depend on the method of their
generation. The power spectrum of the turbulence developed after
injection of a single-mode magnetic field energy peaks at $k_{\rm
max} \sim 3 k_1$, while injection of kinetic energy into an existing
magnetic configuration leads to a turbulent state with power
spectrum peaking at $k_{\rm max} \sim 9 k_1$. Hence, it seems that,
in general, the characteristic length scale of the turbulence is
sensitive to the initial driver: kinetic drivers result in
smaller-scale turbulence states as compared to the case of magnetic
field drivers.

\section{Conclusions}

In this paper we have presented results from direct numerical
simulations of primordial magnetic field evolution during
cosmological phase transitions. These simulations account for the
expansion of the Universe. Simulations were performed to model two
different stages of primordial magnetic field evolution, the first
stage when phase transition processes drive turbulence, and the
second stage when free decay occurs.

We show that different types of initial conditions (drivers) lead
to rapid development of a turbulent state close to equipartition.
 During the following stage we model the free decay of these
turbulence configurations. We study the development during this
free decay stage and analyze characteristic parameters of the
slowly varying power spectrum of the turbulence. We assume that
the properties of the turbulence configurations modeled in our
numerical experiments are similar to those of cosmological
primordial turbulence.

Our simulations allow us to estimate the spectral indices of the
large-scale distribution of the kinetic and magnetic field energies
in developed turbulence. It seems that the final configuration of
primordial magnetic field depends on the initial conditions that
must be determined by the phase transition physics. On the other
hand, the kinetic energy density of the developed state does not
retain any information about the initial conditions and hence is not
sensitive to the details of the phase transition. The spectral index
of the kinetic energy at large scales can be well approximated as 2
(white noise spectrum), while the magnetic energy density spectral
index ranges between 2 and 4 depending on the initial conditions.
Similar spectral indices for the large-scale magnetic field are well
established for laboratory turbulence, see Ref.\ \cite{textbooks}.
Note that simulations in a finite periodic domain may suffer from
the fact that originally disconnected and causally independent
regions come into causal contact within one box turnover time,
which, based on the scale of the domain, is $(v_0 k_1)^{-1}$. For
the case shown in Fig.~\ref{Evolution}, this time is about
$15\,t_1$, but the effective time can be even shorter owing to the
effects of acoustic and Alfv\'en waves.

Our numerical results allow us to estimate typical time-scales for
decaying free turbulence. It seems that cosmological turbulence
decays slightly slower then classical grid turbulence in
laboratory experiments. Although we took a rather small damping
wavenumber, we cannot expect the establishment of a
Kolmogorov-like spectrum at small scales. We see a fast
decorrelation of turbulence at small scales. We note that the
phenomenological approaches developed in
Refs.~\cite{ktr09,caprini2009} imply a fast Kraichnan-like
decorrelation of turbulence. Consequently, only the large-scale
or peak-scale magnetic field results may have cosmological
significance and contribute observable signatures, such as
gravitational wave generation and/or CMB anisotropy production.

We have considered the case of non-helical turbulence. The presence
of weak initial helicity can significantly change the development of
turbulence. This is because magnetic helicity is a conserved
quantity in the limit of large magnetic Reynolds numbers and can
inverse cascade to larger scales \cite{Frisch,Biskamp}, which could
be cosmologically significant. This process has been confirmed
through numerous simulations \cite{enqvist,axel-1,axel,banerjee}.

Summarizing, we find that the generation of a magnetic field at
phase transition scales will lead to the development of turbulent
motions, and, in the case of the electroweak phase transition, this
turbulence has an observable signature in the form of a
gravitational wave signal.

\acknowledgments We appreciate useful discussions with L.
Kisslinger, A. Kosowsky, K.\ Subramanian, and T.\ Vachaspati.
Computing resources have been provided by the Swedish National
Allocations Committee at the Center for Parallel Computers at the
Royal Institute of Technology in Stockholm and the National
Supercomputer Centers in Link\"oping. We acknowledge partial
support from Georgian National Science Foundation grant GNSF
ST08/4-422, Department of Energy grant DOE DE-FG03-99EP41043,
Swiss National Science Foundation SCOPES grant no. 128040, and
NASA Astrophysics Theory Program grant NNXlOAC85G. This work was
supported in part by the European Research Council under the
AstroDyn Research Project 227952 and the Swedish Research Council
grant 621-2007-4064. T.K.\ acknowledges the ICTP associate
membership program and NORDITA for hospitality where this project
was started during the program on electroweak phase transitions.

\newcommand{\yprl}[3]{, Phys.\ Rev.\ Lett.\ {\bf #2}, #3 (#1).}
\newcommand{\yjfm}[3]{, J. Fluid Mech. {\bf #2}, #3 (#1).}
\newcommand{\ypre}[3]{, Phys.\ Rev.\ E {\bf #2}, #3 (#1).}
\newcommand{\ybook}[3]{, {\em #2}. #3 (#1).}

\end{document}